\documentclass[12pt,a4paper]{article}
\pdfoutput=1
\usepackage{jheppub}
\usepackage{graphicx}
\usepackage{lipsum}
\usepackage{tikz}
\usepackage{tikz-cd}
\usepackage{amsmath}
\usepackage{slashed}
\usepackage[all]{hypcap}
\allowdisplaybreaks

\def\Z{\mathbb{Z}}
\def\mod{\text{mod~}}

\usepackage{xspace}

\newenvironment{nalign}{ 
	\begin{equation}
	\begin{aligned}
}{
	\end{aligned}
	\end{equation}
	\ignorespacesafterend
}

\usepackage{amsmath,amscd,amssymb,amsfonts,amsxtra,mathrsfs,graphics,graphicx,amsthm,epsfig,bm,longtable,float,color,tikz,mathtools,xfrac,footnote,rotating,lscape,stmaryrd,wrapfig}
\usepackage[makeroom]{cancel}
\usepackage{bm}
\usepackage{physics}
\usepackage{caption,subcaption}
\usetikzlibrary{arrows}
\usepackage{empheq}
\usepackage{array}   
\newcolumntype{C}{>{$}c<{$}} 

\def\RPn (#1,#2){
  \fill (#1, #2) circle (3pt);
  \fill (#1, #2+1) circle (3pt);
}

\def\sqtwoL (#1,#2,#3){
  \draw[#3] (#1,#2) .. controls (#1-1,#2+1) .. (#1,#2+2);
}

\def\sqtwoR (#1,#2,#3){
  \draw[#3] (#1,#2) .. controls (#1+1,#2+1) .. (#1,#2+2);
}

\def \sqtwoCR (#1,#2,#3){
   \draw[#3] (#1,#2) .. controls (#1+1,#2+.5) and (#1+1.5,#2+2) .. (#1+2,#2+2);
}

\def \sqtwoCL (#1,#2,#3){
   \draw[#3] (#1,#2) .. controls (#1-1,#2+.5) and (#1-1.5,#2+2)  .. (#1-2,#2+2);
}

\def \sqone (#1,#2,#3){
  \draw[#3] (#1,#2) -- (#1,#2+1);
}

\def\Aone (#1,#2){
\fill (#1, #2) circle (3pt);
\fill (#1, #2+1) circle (3pt);
\fill (#1, #2+2) circle (3pt);
\fill (#1, #2+3) circle (3pt);
\fill (#1+2, #2+3) circle (3pt);
\fill (#1+2, #2+4) circle (3pt);
\fill (#1+2, #2+5) circle (3pt);
\fill (#1+2, #2+6) circle (3pt);
\draw (#1, #2) -- (#1, #2+1);
\draw (#1, #2+2) -- (#1, #2+3);
\draw (#1+2, #2+3) -- (#1 + 2, #2+4);
\draw (#1+2, #2+5) -- (#1+2, #2+6);
\draw (#1, #2) .. controls (#1-1, #2+1) .. (#1, #2+2);
\draw (#1+2, #2+4) .. controls (#1+3, #2+5) .. (#1+2, #2+6);
\draw (#1, #2+1) .. controls (#1+1, #2+1.5) and  (#1+1.5 ,#2+3) .. (#1+2,#2+3);
\draw (#1, #2+2) .. controls (#1+1, #2+2.5) and (#1+1.5, #2+4) .. (#1+2, #2+4);
\draw (#1, #2+3) .. controls (#1+1, #2+3.5) and (#1+1.5, #2+5) .. (#1+2, #2+5);
}

\def\rectangle (#1,#2,#3){   \draw[#3] (#1-0.15,#2-0.15) rectangle (#1+0.15,#2+0.15)}

\def\Eone (#1,#2){
\fill (#1, #2) circle (3pt);
\fill (#1, #2+1) circle (3pt);
\fill (#1, #2+2) circle (3pt);
\fill (#1, #2+3) circle (3pt);
\draw (#1, #2) -- (#1, #2+1);
\draw (#1, #2+2) -- (#1, #2+3);
\draw (#1, #2) .. controls (#1-1, #2+1) .. (#1, #2+2);
\draw (#1, #2+1) .. controls (#1+1, #2+2) .. (#1, #2+3);
}

\def\joker (#1,#2){
  \foreach \y in {#2, #2+1, #2+2, #2+3, #2+4}
           {\fill (#1,\y) circle (3pt);}
           \draw (#1,#2) -- (#1, #2+1);
           \draw (#1,#2+3) -- (#1, #2+4);
           \draw (#1,#2+0) .. controls (#1-1,#2+1) .. (#1, #2+2);
           \draw (#1,#2+2) .. controls (#1-1,#2+3) .. (#1, #2+4);
           \draw (#1,#2+1) .. controls (#1+1,#2+2) .. (#1, #2+3);
}

\def\jokercolor (#1,#2, #3){
  \foreach \y in {#2, #2+1, #2+2, #2+3, #2+4}
           {\fill[#3] (#1,\y) circle (3pt);}
           \draw[#3] (#1,#2) -- (#1, #2+1);
           \draw[#3] (#1,#2+3) -- (#1, #2+4);
           \draw[#3] (#1,#2+0) .. controls (#1-1,#2+1) .. (#1, #2+2);
           \draw[#3] (#1,#2+2) .. controls (#1-1,#2+3) .. (#1, #2+4);
           \draw[#3] (#1,#2+1) .. controls (#1+1,#2+2) .. (#1, #2+3);
}

\def\msopart (#1,#2,#3){
    \fill[#3] (#1,#2) circle (3pt); 
      \fill[#3] (#1, #2+1) circle (3pt);
      \fill[#3] (#1, #2+2) circle (3pt);
      \fill[#3] (#1+2, #2+2) circle (3pt);
      \fill[#3] (#1+2, #2+3) circle (3pt);
      \fill[#3] (#1+2, #2+4) circle (3pt);
      \fill[#3] (#1+2, #2+5) circle (3pt);
    \sqtwoCR(#1,#2, #3);
    \sqtwoCR (#1, #2+1, #3);
    \sqtwoCR (#1, #2+2, #3);
    \sqone (#1+2, #2+2, #3);
    \sqone (#1+2, #2+4, #3);
    \sqtwoR(#1+2, #2+3, #3);
    \sqone (#1, #2+1, #3); }

    \def\amme (#1,#2,#3){
       \fill[#3] (#1,#2) circle (3pt) ;
   \sqtwoR(#1,#2,#3);
      \fill[#3] (#1,#2+2) circle (3pt) ;
         \sqone(#1,#2+2,#3);
      \fill[#3] (#1,#2+3) circle (3pt) ;
   \sqtwoR(#1,#2+3,#3);
   \fill[#3] (#1,#2+5) circle (3pt) ;}
   
    \def\questionupsidedon (#1,#2,#3){
       \fill[#3] (#1,#2) circle (3pt) ;
          \sqtwoR(#1,#2,#3);
                 \fill[#3] (#1,#2+2) circle (3pt) ;
             \sqone(#1,#2+2,#3);
                \fill[#3] (#1,#2+3) circle (3pt) ;
}

\newcommand{\fib}[3]{#1 \longrightarrow #2 \longrightarrow #3}

\newcommand{\incl}{\xhookrightarrow{}}

\def\nn{\nonumber}
\def\SO{SO}
\def\O{O}

\def\SU{SU}

\def\U{U(1)}

\def\Spin{\text{Spin}}
\def\A{\mathcal{A}}
\def\th{\text{Thom}}

\def\UU{U}

\def\dd{\text{d}}

\def\dprime{{\prime\prime}}

\def\ext{\text{Ext}}
\def\sq{\text{~Sq}}

\title{Anomaly interplay in $U(2)$ gauge theories}

\author{Joe Davighi}
\author{and Nakarin Lohitsiri}

\affiliation{Department of Applied Mathematics and Theoretical Physics, University of Cambridge, Wilberforce Road, Cambridge, UK}

\emailAdd{jed60@cam.ac.uk}
\emailAdd{nl313@cam.ac.uk}

\abstract{We discuss anomaly cancellation in $U(2)$ gauge theories in four dimensions. For a $U(2)$ gauge theory defined with a spin structure, the vanishing of the bordism group $\Omega_5^{\text{Spin}}(BU(2))$ implies that there can be no global anomalies, in contrast to the related case of an $SU(2)$ gauge theory. We show explicitly that the familiar $SU(2)$ global anomaly is replaced by a local anomaly when $SU(2)$ is embedded in $U(2)$. There must be an even number of fermions with isospin $2r+1/2$, for $r\in \Z_{\geq 0}$, for this local anomaly to cancel. The case of a $U(2)$ theory defined without a choice of spin structure but rather using a spin-$U(2)$ structure, which is possible when all fermions (bosons) have half-integer (integer) isospin and odd (even) $U(1)$ charge, is more subtle. We find that the recently-discovered `new $SU(2)$ global anomaly' is also equivalent, though only at the level of the partition function, to a perturbative anomaly in the $U(2)$ theory, which is this time a combination of a mixed gauge anomaly with a gauge-gravity anomaly. This perturbative anomaly vanishes if there is an even number of fermions with isospin $4r+3/2$, for $r\in \Z_{\geq 0}$, recovering the condition for cancelling the new $SU(2)$ anomaly. Alternatively, this perturbative anomaly can be cancelled by a Wess--Zumino term, leaving a low-energy theory with a global anomaly, which can itself be cancelled by coupling to topological degrees of freedom.
}

\begin{document}
\maketitle
\flushbottom

\section{Introduction}

An $SU(2)$ chiral gauge theory in four dimensions suffers from a non-perturbative global anomaly when there is an odd number of fermion multiplets in isospin $2r+1/2$ representations, for $r\in \Z_{\geq 0}$ ~\cite{Witten:1982fp}. Such a theory is anomalous because the (Euclidean) partition function changes sign under an $SU(2)$ gauge transformation that corresponds to the non-trivial element in $\pi_4(SU(2))=\Z/2$. Equivalently, the anomaly can be seen from a constant gauge transformation by the central element $-{\bf 1}\in SU(2)$, in the  background of a single instanton, as we review in \S \ref{sec:SU2}. 

One might be forgiven for guessing that a $U(2)$ chiral gauge theory suffers from a similar global anomaly, given that $\pi_4(U(2))=\Z/2$ also, and given that $U(2)$ is locally equivalent to $SU(2)\times U(1)$ which has a global anomaly associated with the $SU(2)$ factor. It turns out that this is not the case. A quick way of reaching this conclusion is to recall that global anomalies are detected by the exponentiated $\eta$-invariant~\cite{Witten:1985xe,Dai:1994kq},\footnote{Here we refer to the $\eta$-invariant of an extension of the Dirac operator $i\slashed{D}$ to a five-manifold that bounds spacetime. The $\eta$-invariant of a Dirac operator is a regularized sum of its positive eigenvalues minus its negative eigenvalues, as introduced by Atiyah, Patodi, and Singer~\cite{atiyah_patodi_singer_1976}.} which becomes a bordism invariant when perturbative anomalies vanish. Because the spin-bordism group
\begin{equation}
    \Omega_5^{\text{Spin}}(BU(2))=0  \label{eq1:spinbordismU2}
\end{equation}
(which can be straightforwardly adapted from calculations in~\cite{Davighi:2019rcd,Wan:2019fxh}), the exponentiated $\eta$-invariant must be trivial on all closed spin five-manifolds equipped with a $U(2)$ gauge bundle, which means that there can be no global anomalies in the 4d $U(2)$ gauge theory when perturbative anomalies cancel. In contrast $\Omega_5^{\text{Spin}}(BSU(2))=\Z/2$, which allows for a possible global anomaly in the $SU(2)$ theory.
 
In this paper, our first goal is to explain why there is no global anomaly in a $U(2)$ gauge theory, defined with a choice of spin structure. This is the subject of \S \ref{sec:U2}. The argument is simple enough to summarise in this Introduction. Recall firstly that $U(2)$ may be written as
\begin{equation}
    U(2) \cong \frac{SU(2)\times U(1)}{\Z/2},
\end{equation}
where the $\Z/2$ quotient is generated by the central element $(-\mathbf{1},e^{i\pi})\in SU(2)\times U(1)$. As for the $SU(2)$ case, one could make a constant gauge transformation by the element $(-\mathbf{1},1)\in SU(2)\times U(1)$ in the background of a single instanton, and might thus be tempted to reach the same conclusion that there can be a global anomaly. However, this gauge transformation is equivalently described by the element $(\mathbf{1},e^{i\pi})\in SU(2)\times U(1)$. Thus, the anomalous transformation is in fact a local $U(1)$ transformation, and we can compute the variation of the fermionic partition function using the appropriate counterterms in the effective action. The non-invariance of the path integral measure (when there is an odd number of multiplets with isospin $2r+1/2$) arises simply because there is a mixed triangle anomaly. 

We show explicitly that the (perturbative) mixed triangle anomaly can vanish only if there is an even number of multiplets with isospin $2r+1/2$, by reducing the anomaly cancellation condition modulo 2.\footnote{In \S \ref{sec:bordism} we arrive at the same conclusion by directly computing the $\eta$-invariant using the Atiyah--Patodi--Singer (APS) index theorem~\cite{atiyah_patodi_singer_1976}.} Note that this is only true when the global structure of the gauge group is strictly $U(2)$. The argument does not follow for the (locally isomorphic) gauge group $SU(2)\times U(1)$, even though the formula for the perturbative anomaly is the same, because not every representation of $SU(2)\times U(1)$ corresponds to a representation of $U(2)$. Having realised that the apparently global $SU(2)$ anomaly is manifest in $U(2)$ rather as a local anomaly, we may conclude from \eqref{eq1:spinbordismU2} that there can be no other new global anomalies in a $U(2)$ theory (defined with a spin structure).

Understanding the absence of global anomalies in a $U(2)$ gauge theory, but nonetheless the necessity of the condition on isospin $2r+1/2$ multiplets, is of some phenomenological interest, because $U(2)$ could be the gauge group for the electroweak theory~\cite{Tong:2017oea}. For example, anomaly cancellation in such a theory provides constraints on the electroweak quantum numbers of field content in the context of going beyond the Standard Model.

We then turn to the more subtle case of a $U(2)$ gauge theory defined without a spin or spin$_c$ structure, and perform a similar analysis relating to the `new $SU(2)$ (global) anomaly' that afflicts an $SU(2)$ gauge theory that is similarly defined without a spin structure~\cite{Wang:2018qoy}. Recall that fields in such a theory are instead defined using a spin-$SU(2)$ structure, which requires that all fermions (bosons) have half-integer (integer) isospin. The $SU(2)$ theory is anomalous if there is an odd number of fermion multiplets with isospin $4r+3/2$, for $r\in \Z_{\geq 0}$. The partition function for such a theory, defined on certain manifolds that are not spin (in particular, on $\mathbb{C}P^2$), changes sign under the combined action of a diffeomorphism $\varphi$ and an $SU(2)$ gauge transformation $W$. This is the new $SU(2)$ anomaly, which we shall recap in \S \ref{sec:SU2}.

The second goal of this paper is to understand what happens to the new $SU(2)$ anomaly in the analogous situation in which the gauge group is enlarged from $SU(2)$ to $U(2)$. If the field content is such that all fermions (bosons) have half-integer (integer) isospins and odd (even) $U(1)$ charges, then the $U(2)$ gauge theory can be defined without a spin structure, using this time a spin-$U(2)$ structure to parallel transport fields. Again, one might expect that a global anomaly should afflict such a theory, corresponding to the new $SU(2)$ anomaly; and again, this turns out not to be the case, as we show in \S \ref{sec:no spin structure}.

The new $SU(2)$ anomaly enjoys a similar but subtly different fate to the old one. This time, because of the crucial role played by the diffeomorphism $\varphi$ in deriving the new $SU(2)$ anomaly, we find that the anomalous combination of $\varphi$ and $W$ cannot be replaced by a local $U(2)$ gauge transformation, as was the case for the `old' $SU(2)$ anomaly. However, the anomalous combined action of $\varphi$ and $W$ has the same effect on the fermionic partition function as a local $U(2)$ gauge transformation with determinant $-1$. This gives rise to a local anomaly, that is a combination of the mixed triangle anomaly (corresponding to a Feynman diagram with two external $SU(2)$ currents and one $U(1)$ current) with the gauge-gravity anomaly for the $U(1)$ current. By considering this particular combination of perturbative anomalies reduced modulo 4, we find that the $U(2)$ gauge theory defined using a spin-$U(2)$ structure can only be anomaly-free when there is an even number of fermion multiplets with isospin $4r+3/2$.

It is important to stress that, in the $U(2)$ theory, this condition on isospin $4r+3/2$ multiplets must be satisfied simply for perturbative anomalies to cancel; thus, unlike the new $SU(2)$ anomaly, this condition persists even if we choose to restrict our attention to spin manifolds.

In \S \ref{sec:SU2} we review the pair of global anomalies in $SU(2)$ gauge theory. In \S \ref{sec:U2} we discuss the $U(2)$ theory defined using a spin structure, before turning to the case without spin structure in \S \ref{sec:no spin structure}.   Finally, in \S \ref{sec:bordism} we interpret our results in terms of cobordism invariants. We thence explain why there are no other global anomalies in the $U(2)$ theory defined using a spin-$U(2)$ structure.

\section{Review of the \texorpdfstring{$SU(2)$}{SU(2)} global anomalies} \label{sec:SU2}

\subsection*{The old anomaly}

We first review the global anomaly that occurs for an $SU(2)$ gauge theory defined on a four-manifold $M$ (which we take to be Euclidean) using a spin structure~\cite{Witten:1982fp}. Consider a single fermion transforming in the isospin-$j$ representation, coupled to a background $SU(2)$ gauge field with curvature $F$. Let $n_+$ ($n_-$) denote the number of fermion modes with positive (negative) chirality ({\em i.e.} eigenvalue under $\gamma^5$). The
Atiyah--Singer index theorem tells us that
\begin{equation}
n_+-n_- = -\frac{1}{8\pi^2}\int_M \text{Tr~} F \wedge F = -T(j)\  p_1(F), \label{eq:Atiyah-Singer}
\end{equation}
where $p_1(F)\in \Z$ is the first Pontryagin number (or instanton number), and 
\begin{equation}
    T(j)=\frac{2}{3}j(j+1)(2j+1) \label{eq:Dynkin}
\end{equation}
is the Dynkin index
 defined via $\text{Tr}(t^a_j t^b_j)=\frac{1}{2}T(j) \delta^{ab}$. Here $\{t^a_j\}$ denotes a basis for the isospin-$j$ representation of $\mathfrak{su}(2)$. Because $n_+-n_-$ is congruent to $n_+ +n_-\equiv \mathcal{N}_j $ modulo 2, the total number of fermion zero modes satisfies
\begin{equation} \label{eq:0modes}
    \mathcal{N}_j \equiv T(j)\ p_1(F) \quad (\mod 2).
\end{equation}
If $\mathcal{N}_j$ is odd, then the partition function will change sign under the action of $(-1)^F$, where $F$ is the fermion number. But since $(-1)^F$ is equivalent to applying a gauge transformation by the central element $-\mathbf{1}\in SU(2)$, this implies that $SU(2)$ is anomalous in such a scenario.

Only fermions with isospin $j=2r+1/2$ can contribute to this anomaly, and only in backgrounds with odd instanton number, because it is only for these values of $j$ that the Dynkin index (\ref{eq:Dynkin}) is odd. Thus, the anomaly vanishes if and only if the following holds
\begin{equation} \label{eq:condition 1}
    \begin{minipage}{0.8\textwidth}
\underline{Condition 1:} There is an even number of fermions transforming in representations with isospin $2r+1/2$, for $r\in \Z_{\geq 0}$.     
    \end{minipage}
\end{equation}
This is the familiar $SU(2)$ anomaly discovered by Witten~\cite{Witten:1982fp}.

\subsection*{The new anomaly}

Suppose now that there is no spin structure available, and that fermions are instead defined using a weaker spin-$SU(2)$ structure.\footnote{The idea of using such `spin-$G$' structures, for various Lie groups $G$ (going beyond the case where $G=U(1)$), was introduced in Refs.~\cite{Avis:1979de,Back:1978zf}. } The transition functions for a spin-$SU(2)$ bundle are valued in the group
\begin{equation}
    \text{Spin}_{SU(2)}(4) \equiv \frac{\text{Spin}(4)\times SU(2)}{\Z/2},
\end{equation}
where the $\Z/2$ quotient is generated by the central element $-\mathbf{1}$ of $SU(2)$ paired with the element $(-1)^F\in\text{Spin}(4)$. All fields must transform in representations of this group, which requires that all fermions have half-integer isospin, and all bosons have integer isospin. Such a theory can be defined on all orientable four-manifolds, including those that are not spin such as $\mathbb{C}P^2$.\footnote{It was first observed that a fermionic theory can be defined on $\mathbb{C}P^2$, using a spin$_c$ structure, in Ref.~\cite{Hawking:1977ab}.
Indeed, every orientable four-manifold admits a spin$_c$ structure -- but one must assume that $M$ is equipped with a spin-$SU(2)$ structure, and not a spin$_c$ structure, in order to see the new $SU(2)$ anomaly.}

In the simpler case that we discussed above, we saw how the usual $SU(2)$ anomaly could be seen from the action of $(-1)^F$ on the path integral measure, since $(-1)^F$
is equivalent to an $SU(2)$ gauge transformation by $-\mathbf{1}\in SU(2)$. The new $SU(2)$ anomaly is more subtle, and cannot be seen from a pure gauge transformation. Rather, the new $SU(2)$ anomaly is the non-invariance of the path integral under a transformation $\hat{\varphi}$ which is a combined diffeomorphism $\varphi$ of $M$ (for certain non-spin manifolds $M$) with an $SU(2)$ gauge transformation $W$.

To see this anomaly one may take $M$ to be $\mathbb{C}P^2$, and 
$\varphi:z_i\mapsto z_i^*$
to act by complex conjugation on the homogeneous complex coordinates $\{z_i\}$ of $\mathbb{C}P^2$. A spin-$SU(2)$ connection $A$ may be defined by embedding a spin$_c$ connection $a$ in $\mathfrak{su}(2)$, {\em viz.} $A=\sigma^3 a$, where $\sigma^3$ is the diagonal Pauli matrix. The spin$_c$ connection $a$ obeys the following quantisation condition
\begin{equation}
    \int_S \frac{da}{2\pi} \equiv \frac{1}{2}\int_S w_2(TM) \quad (\mod 1), \label{eq:SW class constraint}
\end{equation}
for any closed oriented 2-manifold $S\subset M$, where $w_2(TM)$ is the second Stiefel--Whitney class, which is such that  $2a$ defines a properly-normalised $U(1)$ gauge field. In particular, choose a spin$_c$ connection $a$ such that 
\begin{equation}
    \int_{\mathbb{C}P^1} \frac{da}{2\pi} = \frac{1}{2} \label{eq: spinc normalisation}
\end{equation}
for some $\mathbb{C}P^1 \subset \mathbb{C}P^2$. Such a spin$_c$ connection reverses sign under the diffeomorphism $\varphi$. The spin-$SU(2)$ connection $A$, however, is invariant under the combined action of $\varphi$ with any $SU(2)$ gauge transformation $W$ which also flips its sign, such as $W=\begin{psmallmatrix}
    0 & -1\\
    1 & 0
    \end{psmallmatrix}$.

An anomaly in the transformation $\hat{\varphi}$ has to arise from the path integral over the fermion zero modes. 
On $\mathbb{C}P^2$ the number of zero modes $\mathcal{N}_j$ equals the index of the Dirac operator $\mathfrak{J}_j$ (they are not only congruent modulo 2 as before).\footnote{This is because on $\mathbb{C}P^2$ the Dirac operator only has zero modes of one chirality.
}
For a single fermion multiplet in the isospin-$j$ representation coupled to the background spin-$SU(2)$ connection $A$ defined above, the Atiyah--Singer index theorem implies the index is~\cite{Wang:2018qoy}
\begin{equation}\label{eq:0modes2}
    \mathfrak{J}_j=\mathcal{N}_j = \frac{1}{24}(4j^2-1)(2j+3).
\end{equation}
The zero modes come in pairs with eigenvalues $+1$ and $-1$ under $\hat{\varphi}$. Hence, the fermionic partition function $Z[A]$ transforms under the action of $\hat{\varphi}$ by 
\begin{equation} \label{eq:phihat_on_Z}
    Z[A] \xrightarrow{\hat{\varphi}} (-1)^{\mathfrak{J}_j/2} Z[A].
\end{equation}
The index $\mathfrak{J}_j$ is even for all half-integer values of $j$, but is congruent to 2 mod 4 only when $j=4r+3/2$ for $r\in \Z_{\geq 0}$. For all other half-integer values of $j$, the index $\mathfrak{J}_j$ is divisible by 4. Hence, the partition function is invariant under $\hat{\varphi}$, and the theory is non-anomalous, if and only if the following condition holds:
\begin{equation} \label{eq:condition 2}
    \begin{minipage}{0.8\textwidth}
\underline{Condition 2:} there is an even number of fermions transforming in representations with isospin $4r+3/2$, for $r\in \Z_{\geq 0}$.     
    \end{minipage}
\end{equation}
This is the new $SU(2)$ anomaly recently discovered by Wang, Wen, and Witten~\cite{Wang:2018qoy}.

\section{\texorpdfstring{$U(2)$}{U(2)} gauge theory with a spin structure} \label{sec:U2}

We now turn to $U(2)$ gauge theory. We begin with the simpler case of a $U(2)$ gauge theory defined with a spin structure, for which the vanishing of the bordism group \eqref{eq1:spinbordismU2} implies there are no global anomalies. We will here give a physical explanation of this fact, previously noted in Refs.~\cite{Davighi:2019rcd,Wan:2019fxh}, which demonstrates the subtle interplay between local and global anomalies in $U(2)$.

The representation theory of $U(2)$ plays a crucial role in the arguments used in this paper. Recall that an irreducible representation of $U(2)\cong (SU(2)\times U(1))/\Z/2$ is labelled an irreducible representation of $SU(2)$, itself labelled by an isospin $j$, together with a $U(1)$ charge $q$, subject to a restriction relating $q$ and $j$. Namely, $q$ and $j$ must satisfy the following `isospin-charge relation'\footnote{We note in passing that this isospin-charge relation \eqref{eq: charge constraint} is satisfied by all the SM fermion fields, where $U(1)$ corresponds to hypercharge. Hence the electroweak gauge symmetry could be either $SU(2)\times U(1)$ or $U(2)$. }
\begin{equation}
q \equiv 2j \quad (\mod 2), \label{eq: charge constraint}
\end{equation}
in convenient units where both gauge couplings are set to one.

Consider a theory with a single fermion with isospin $j$ and charge $q$ (satisfying \eqref{eq: charge constraint}), coupled to a background $U(2)$ gauge field with curvature $F$ and defined on $S^4$.
Recall that the usual $SU(2)$ anomaly occurs when the fermionic partition function changes sign under the gauge transformation by $-\mathbf{1}\in SU(2)$. Embedding $SU(2)\subset U(2)$, this global $SU(2)$ transformation is equivalent to a $U(1)$ gauge transformation by $e^{i\pi}$, which is a local gauge transformation. 

The variation of the partition function $Z[A]$
under a potentially anomalous $U(1)$ gauge transformation can be computed using the appropriate counterterms in the effective action (see {\em e.g.}~\cite{Preskill:1990fr}).
For a $U(1)$ transformation by angle $\theta$, we have that
\begin{nalign}
    Z[A] &\rightarrow Z[A] \exp \left[ -\frac{iq\theta}{8 \pi^2} \int_{S^4} \text{Tr~} F \wedge F + \text{gravitational piece}
    \right] \\
    &= Z[A] \exp \left[ -iq\theta\ T(j)\ p_1(F) + \text{gravitational piece}
    \right],
\end{nalign}
where the gravitational piece is proportional to the integral of $\text{Tr~} R\wedge R$ which vanishes for $S^4$. Setting $\theta=\pi$ and the instanton number $p_1(F)=1$, this reduces to
\begin{equation}
    Z[A]\rightarrow (-1)^{qT(j)} Z[A].
\end{equation}
We see that the path integral is invariant under this transformation if and only if $q T(j)$ is even.

Recall that the Dynkin index $T(j)$ is only odd for isospins $j\in 2\Z_{\geq 0} +1/2$. The isospin-charge relation \eqref{eq: charge constraint} means that $q$ is also odd for these representations. Hence, there is necessarily an anomaly if there is an odd number of fermions in multiplets with isospin $2r+1/2$; in other words, precisely when condition \eqref{eq:condition 1} is violated.
Thus, we find that the $SU(2)$ global anomaly manifests itself rather as a perturbative anomaly when $SU(2)$ is embedded in $U(2)$. There are no global anomalies in the $U(2)$ theory.

Indeed, one can directly derive that condition \eqref{eq:condition 1} must hold for a $U(2)$ gauge theory by considering the equations for perturbative anomaly cancellation. 
Suppose that we have $N_j$ fermions transforming in isospin-$j$ representations of $U(2)$, with charges $\{q_{j,\alpha}\}$, where $\alpha=1,\dots N_j$. We assume without loss of generality that all fermions have left-handed chirality. The mixed triangle anomaly (that is, the triangle anomaly involving two $SU(2)$ gauge bosons and one $U(1)$ gauge boson) is proportional to
\begin{equation} \label{eq:SU22 U1}
 \mathcal{A}_{\text{mix}} \equiv \sum_j T(j) \sum_{\alpha=1}^{N_j} q_{j,\alpha}=0,    
\end{equation}
\textit{}The fact that $T(j)$ is odd only for $j\in 2\Z_{\geq 0}+1/2$, together with the isospin-charge relation, means that reducing mod 2 immediately yields
\begin{equation} \label{eq: no witten anomaly}
\sum_{j\in 2\Z+1/2} 1 \equiv 0 \quad (\mod 2),
\end{equation}
and hence that condition \eqref{eq:condition 1} must be satisfied to avoid a perturbative mixed anomaly. It is possible to give a unified discussion of the perturbative and non-perturbative anomalies in this theory by computing the $\eta$-invariant explicitly. We give such an account in \S \ref{sec:bordism}.

\section{\texorpdfstring{$U(2)$}{U(2)} gauge theory without a spin structure} \label{sec:no spin structure}

We now turn to the case where a spin structure is not available. Instead, we can use a spin-$U(2)$ structure to parallel transport fields, provided that all fields transform in representations of the group 
\begin{equation}
    \text{Spin}_{U(2)} \equiv \frac{\text{Spin}(4)\times U(2)}{\Z/2}.
\end{equation}
The $\Z/2$ quotient is generated by the product of the element $(-1)^F\in \text{Spin}(4)$ with the central element $-\mathbf{1}\in U(2)$. Recalling also the effects of the $\Z/2$ quotient within $U(2)$, we have the following constraints on the allowed representations:
\begin{nalign}
    \text{fermion~} &\longleftrightarrow j \in (2\Z+1)/2  &&\longleftrightarrow q \text{~odd}, \\
    \text{boson~} &\longleftrightarrow \qquad j\in \Z  &&\longleftrightarrow q \text{~even},
\end{nalign}
where $(q,j)$ label the $U(2)$ representations as before. 

In the analogous $SU(2)$ theory, the new $SU(2)$ anomaly is associated with a transformation $\hat{\varphi}$ that is a combined diffeomorphism $\varphi$ plus gauge transformation $W$, as we reviewed in \S \ref{sec:SU2}. Recall that $\hat{\varphi}$ acts on the partition function as
\begin{equation}
    Z[A] \xrightarrow{\hat{\varphi}} (-1)^{\mathfrak{J}_j/2} Z[A]. \label{eq:phi hat on Z}
\end{equation}
Let us first analyse the behaviour of the $U(2)$ theory under this same transformation.
To that end, again take $M$ to be $\mathbb{C}P^2$, and as in \S \ref{sec:SU2} define $\hat{\varphi}$ to be the combination of the complex conjugation diffeomorphism $\varphi:z_i \mapsto z_i^*$ with the $U(2)$ gauge transformation $W=\begin{psmallmatrix}
    0 & -1\\
    1 & 0
    \end{psmallmatrix}$. 
Moreover, we define a spin-$U(2)$ connection $A=\sigma^3 a$, where $a$ is the spin$_c$ connection satisfying  Eqs. (\ref{eq:SW class constraint}, \ref{eq: spinc normalisation}), which is invariant under $\hat{\varphi}$. 

The diffeomorphism $\varphi$ (on its own) is such that $\varphi^2=-1$ when acting on fermions. More specifically, $\varphi$ can be thought of as a certain spatial rotation through an angle $\pi$, corresponding (in certain coordinates) to the following transformation on a 2-component Weyl fermion $\psi_a$:
\begin{equation}
    \psi_{a} \xmapsto{\varphi} 
    \begin{pmatrix}
    i & 0\\
    0 & -i
    \end{pmatrix}
    \psi_{a}, \label{eq:spatial rotation}
\end{equation}
where the index labels Lorentz $SU(2)$ indices of the spin-$1/2$ fermion.  Because the matrix appearing in \eqref{eq:spatial rotation} is not proportional to the identity,
this diffeomorphism cannot therefore be subsumed by the $U(1)$ phase degree of freedom in $U(2)$. Thus, as in the $SU(2)$ case, the transformation $\hat{\varphi}$ is necessarily not equivalent to a pure $U(2)$ gauge transformation. Since $\hat{\varphi}$ is inequivalent to a local gauge transformation, in contrast to the situation in \S \ref{sec:U2}, we might suspect that this new $SU(2)$ global anomaly will stick around in the $U(2)$ theory.

However, what we can do instead is construct a local $U(2)$ gauge transformation whose action on the fermionic partition function $Z[A]$ is identical to \eqref{eq:phi hat on Z}. Consequently, cancellation of perturbative anomalies shall guarantee that the suspected global anomaly in fact vanishes. To wit, consider a gauge transformation by
\begin{equation} \label{eq:U1gt}
    \Tilde{W}(\theta) = 
    \begin{pmatrix}
    e^{i\theta} & 0\\
    0 & e^{i\theta}
    \end{pmatrix} \in U(2), \qquad \theta \notin \pi \Z, 
\end{equation}
{\em i.e.} by a pure $U(1)$ phase. Note that $\text{det}\ \Tilde{W}\neq 1$ for $\theta \notin \pi \Z$, so that there is no corresponding gauge transformation in $SU(2)$ by design.
Let us now compute the transformation of $Z[A]$ under $\Tilde{W}(\theta)$, for a single fermion multiplet with isospin-$j$ and charge $q$ coupled to the spin-$U(2)$ connection $A$. This time the gravitational contribution will be non-vanishing because $\mathbb{C}P^2$ has non-zero signature. Taking into account the contributions from both the mixed gauge anomaly and the gauge-gravity anomaly, the shift in the Euclidean partition function, for now on a general four-manifold $M$ with metric $g$, is
\begin{equation}
        Z[A] \rightarrow Z[A] \exp \left(  -S_{\text{gauge}} -S_{\text{grav}}    \right),
\end{equation}
where 
\begin{equation}
    S_{\text{gauge}}=-\frac{i\theta}{16 \pi^2} q\int_{M} \text{Tr~} F_{\mu\nu} \tilde{F}^{\mu\nu} d^4x, 
    \label{eq:gauge action}
\end{equation}
in which the trace is only over the $SU(2)$ gauge indices (we here choose to keep Lorentz indices explicit for clarity), and 
\begin{equation}
    S_{\text{grav}}=-\frac{i\theta\ }{16\pi^2}\frac{\text{Tr}(Q)}{24}   \int_{M}  R_{\mu\nu\sigma\tau} \tilde{R}^{\mu\nu\sigma\tau}\sqrt{g}\; d^4x,
    \label{eq:gravity action}
\end{equation}
where $Q$ is the generator of the $U(1)$ factor in $U(2)$, and the trace sums over all $2j+1$ components of the isospin-$j$ representation. Recall that $\tilde{F}^{\mu\nu} = \frac{1}{2}\epsilon^{\mu\nu\sigma\tau}F_{\sigma\tau}$ and $\tilde{R}^{\mu\nu\sigma\tau}=\frac{1}{2}\epsilon^{\mu\nu\alpha\beta} R_{\alpha\beta}^{\ \ \ \sigma\tau}$, where $R_{\mu\nu\sigma\tau}$ are the components of the Riemann tensor.

We can relate both these integrals to characteristic classes of bundles over $M$, taking care with the various normalisation factors. Noting that $\tau^a=\sigma^a/2$ are the generators of the $SU(2)$ factor of $U(2)$, the choice $A=\sigma^3 a$ implies that $F^a_{\mu\nu}=2\delta^{a3}f_{\mu\nu}$, where $f=da$ is the curvature of the spin$_c$ connection $a$. We can thus reduce \eqref{eq:gauge action} to an integral over the spin$_c$ connection,
\begin{equation}
    S_{\text{gauge}}=-\frac{iq\theta}{4 \pi^2} \left(\frac{T(j)}{2} \right) \int_{M} f_{\mu\nu} \tilde{f}^{\mu\nu} d^4x = 
    -\frac{iq\theta}{4 \pi^2} T(j) \int_{M} f\wedge f.
\end{equation}
The normalisation \eqref{eq: spinc normalisation} of the spin$_c$ connection determines its first Pontryagin class in terms of the signature $\sigma$ of $M$, {\em viz.}
\begin{equation}
    \frac{1}{2} \int_M \frac{f\wedge f}{(2\pi)^2} = \frac{1}{8}\sigma.
\end{equation}
Since $\sigma=1$ for $\mathbb{C}P^2$ we have that, when $M=\mathbb{C}P^2$, 
\begin{equation}
    S_{\text{gauge}}=-\frac{i\theta}{4} T(j)q.
\end{equation}
For the gravitational contribution, we use the fact that
\begin{equation}
    -\frac{1}{16\pi^2}\int_{M} R_{\mu\nu\sigma\tau} \tilde{R}^{\mu\nu\sigma\tau} \sqrt{g}  d^4x = \frac{1}{2} \int_{M} \frac{\text{Tr~} R\wedge R}{(2\pi)^2}=p_1[M]=3\sigma(M),
\end{equation}
and that $\text{Tr}(Q)=(2j+1)q$ to deduce that
\begin{equation}
    S_{\text{gravity}}=+\frac{i\theta}{8} (2j+1)q
\end{equation}
when $M=\mathbb{C}P^2$.

The partition function therefore shifts by
\begin{equation}
        Z[A] \rightarrow Z[A] \exp \left[ -\frac{i\theta}{4} \left(T(j)-\frac{1}{2}(2j+1)
        \right) q
    \right].
\end{equation}
Using the expression \eqref{eq:Dynkin} for the Dynkin index, we find that the factor in square brackets is nothing but $-i\theta \mathfrak{J}_j q$,
where $\mathfrak{J}_j$ is the same index from \eqref{eq:0modes2} that detected the new $SU(2)$ anomaly. Therefore, setting $\theta=\pi/2$ gives
\begin{equation}
        Z[A] \xrightarrow{\Tilde{W}(\pi/2)} (-1)^{\mathfrak{J}_j q/2} Z[A].
\end{equation}
Recalling that all fermions in this theory have half-integral isospin $j$ and odd charge $q$, and that $T(j)\equiv 2 \; (\mod 4)$ only when $j\in 4\Z +3/2$, we see that there is a perturbative $U(2)$ anomaly when there is an odd number of fermion multiplets with isospin $j\in 4\Z_{\geq 0} +3/2$; in other words, precisely when condition \eqref{eq:condition 2} is violated.

Another way to see that the $U(2)$ gauge transformation by $\Tilde{W}(\pi/2)$ has the same action on the path integral as the action $\hat{\varphi}$ of the diffeomorphism $\varphi$ plus $SU(2)$ gauge transformation $W$ is to consider the composition $\hat{\varphi}(\pi/2)\equiv\hat{\varphi}\cdot \Tilde{W}(\pi/2)$ of these two transformations. In other words, consider the combined action on $Z[A]$ of the diffeomorphism $\varphi$ plus a $U(2)$ gauge transformation by $\Tilde{W}(\pi/2)\cdot W = iW$. The argument proceeds almost exactly as the argument for the new $SU(2)$ anomaly, as summarised in \S \ref{sec:SU2}; the only difference is that now the fermion zero modes transform in pairs under $\hat{\varphi}(\pi/2)$ with eigenvalues $+i$ and $-i$ (rather than $+1$ and $-1$) whose product is now $+1$ (rather than $-1$ as before). Thus, since there is an even number of zero modes, the action of $\hat{\varphi}\cdot \Tilde{W}(\pi/2)$ is always non-anomalous, and so each of $\hat{\varphi}$ and $\Tilde{W}(\pi/2)$ must contribute the same mod 2 anomaly.

As we saw in \S \ref{sec:U2} for the old $SU(2)$ anomaly, we can again deduce the necessity of condition \eqref{eq:condition 2} directly from the equations for perturbative anomaly cancellation. This time, however, we also need to use the cancellation of the gauge-gravity anomaly, 
\begin{equation}
 \mathcal{A}_{\text{grav}} \equiv \sum_j (2j+1) \sum_{\alpha=1}^{N_j}  q_{j,\alpha}=0. \label{eq: gravity anomaly}
\end{equation}
If we take a particular linear combination of local anomaly equations, {\em viz.} $\frac{1}{4}$[(\ref{eq:SU22 U1})-$\frac{1}{2}($\ref{eq: gravity anomaly})], we obtain
 \begin{equation}
 \sum_{j \text{~half integer}} \mathfrak{J}_j \sum_\alpha q_{j,\alpha} = 0. \label{eq: lin com}
 \end{equation}
Reducing this equation modulo 4, and using the properties of $\mathfrak{J}_j$ noted above, we immediately obtain
 \begin{equation} \label{eq: no WWW anomaly}
\sum_{j=4r+3/2} 1 \equiv 0 \quad (\mod 2),
\end{equation}
recovering the condition \eqref{eq:condition 2} that, in the $SU(2)$ case, is required to cancel the new $SU(2)$ anomaly.

\subsection{Interpretation of the $U(2)$ anomalies}

We have now seen how both conditions \eqref{eq:condition 1} and \eqref{eq:condition 2}, for the cancellation of the old and new $SU(2)$ anomalies, do not correspond to global anomalies when $SU(2)$ is embedded as a subgroup of $U(2)$. The arguments used for the two anomalies were, however, qualitatively different. In the case of the old $SU(2)$ anomaly, for a theory defined using a spin structure, the global transformation in $SU(2)$ corresponds to a local transformation in $U(2)$, for which there is an associated perturbative anomaly if there are an odd number of multiplets with isospin $j\in 2\Z_{\geq 0}+1/2$. 

For the new $SU(2)$ anomaly, however, the mixed diffeomorphism plus gauge transformation is not equivalent to a local transformation in $U(2)$. It nonetheless transpires to be equivalent to a local transformation in $U(2)$ at the level of its action on the fermionic partition function. In this sense, the condition \eqref{eq:condition 2} emerges somewhat coincidentally from perturbative anomaly cancellation in the $U(2)$ theory, which should be thought of as `trivialising' the new $SU(2)$ global anomaly; for the old $SU(2)$ anomaly, the correct interpretation is rather that there is no global anomaly at all in $U(2)$.

As a result, the condition \eqref{eq:condition 2} enjoys a different `status' in the $SU(2)$ theory versus the $U(2)$ theory.
It is important to recall that the new $SU(2)$ anomaly is no barrier to the consistency of an $SU(2)$ gauge theory when formulated only on spin manifolds.\footnote{In fact, the new $SU(2)$ anomaly is not an insurmountable barrier to consistency on non-spin manifolds either; in this case, one can couple to a topological quantum field theory (tQFT), in the same 4d bulk, which has the same anomaly theory (specifically, this anomaly theory has 5-form lagrangian given by the product $w_2 w_3$ of Stiefel--Whitney classes), and thereby cancel the $\Z/2$-valued global anomaly. This kind of anomaly cancellation mechanism was introduced as a `topological Green--Schwartz mechanism' in~\cite{Garcia-Etxebarria:2017crf}. Note that the tQFT to which we couple has no propagating degrees of freedom that would alter the phenomenology of the theory. } In contrast, the constraint \eqref{eq: no WWW anomaly} on the $U(2)$ theory is required by $U(2)$ gauge invariance, and so its violation, like the violation of the original Witten anomaly, would render the $U(2)$ theory inconsistent (even on spin manifolds). 
  
\subsection{Disentangling the anomaly interplay} \label{sec:disentangle}

It is possible to make rigorous the claim that the condition \eqref{eq:condition 2} emerges only coincidentally in the $U(2)$ theory without spin structure. In fact, in this Section we show that, at least at the level of effective field theory, the perturbative anomaly may be cancelled to leave behind a theory with the `new' type of global anomaly, thereby disentangling the anomaly interplay described above.

For instance, if one interprets the $U(2)$ gauge theory described in Section \ref{sec:no spin structure} as an effective field theory of the light excitations that is valid only up to some momentum cutoff scale $\Lambda$, then Wess--Zumino (WZ) terms may be included in the lagrangian which cancel anomalies in the low-energy theory.\footnote{The mechanism we describe here for cancelling anomalies at low-energies might also be referred to as a `Green--Schwartz mechanism'~\cite{Green:1984sg}, a terminology that stems from a famous application to cancelling mixed anomalies in string theory.} If we consider again a general spectrum with $N_j$ fermions transforming with isospin-$j$ and with charges $\{q_{j,\alpha}\}$, then let us modify the effective lagrangian by adding the pair of WZ terms~\cite{Preskill:1990fr,Wess:1971yu,AlvarezGaume:1983ig}
\begin{equation} \label{eq:WZ}
\mathcal{L} \to \mathcal{L} + \mathcal{L}_{\text{WZ}}, \qquad \mathcal{L}_{\text{WZ}}=\frac{i\mathcal{A}_{\text{mix}}}{32\pi^2} \phi F_{\mu\nu}^a \tilde{F}^{a \mu\nu} +\frac{i \mathcal{A}_{\text{grav}}}{384\pi^2} \phi \sqrt{g}R_{\mu\nu\sigma\tau} \tilde{R}^{\mu\nu\sigma\tau},
\end{equation}
where $\phi(x)$ is a dimensionless (circle-valued) pseudoscalar field which enjoys a shift symmetry under the $U(1)$ factor in $U(2)$, {\em viz.} $\phi(x) \to \phi(x) + \theta$ for $g=e^{i\theta}$,\footnote{We remark that these WZ terms are well-defined even though $\phi$ is circle-valued; under the `large gauge transformation' $\phi(x) \to \phi(x) + 2\pi$, the phase of the exponentiated action shifts by an integer multiple of $2\pi$ and so the path integral is unchanged, for any orientable 4-manifold $M$ and for any fermion content.} and is a singlet under the $SU(2)$ part.
These WZ terms conveniently encode the effects of integrating out a ``mirroring'' set of heavy chiral fermions, which transform in the same set of $U(2)$ representations but with opposite chirality.\footnote{We might imagine that heavy masses could arise from Yukawa-like interactions with a Higgs field. However, the precise construction of a suitable Yukawa sector is not immediately obvious, and we do not venture the details of a UV completion here.}

One can check explicitly that under any $U(2)$ gauge transformation, including generic $U(1)$ transformations of the form (\ref{eq:U1gt}), the effective lagrangian is now invariant; the shifts of the WZ terms precisely cancel the shift in the effective action due to the non-invariance of the path integral measure for the chiral fermions, as is the purpose of the construction. However, gauge invariance comes at a price, which is that the full $U(2)$ symmetry is no longer linearly-realised. To see this, note that invariance under local $U(1)$ gauge transformations $\phi(x) \to \phi(x) + \theta(x)$, for a smooth function $\theta(x)$, requires that the pseudoscalar $\phi$ should have a kinetic term of the Stueckelberg form, that is
\begin{equation}
\mathcal{L} \supset \frac{1}{2}|d\phi - b|^2, 
\end{equation}
where $b$ is the $U(1)$ component of the spin-$U(2)$ connection, which transforms as $b \to b+d\theta$.\footnote{Locally, $b$ behaves like a $U(1)$ gauge field.} Thus, the component $b$ becomes massive, meaning that at low-energies only a subgroup $SU(2) \subset U(2)$ is linearly-realised.

Interestingly, adding WZ terms to the effective lagrangian is not guaranteed to cancel the more subtle global anomalies. In the presence of the WZ terms (\ref{eq:WZ}), one may now consider fermion content which violates condition (\ref{eq:condition 2}) without violating perturbative anomaly cancellation. For such a theory, we should reconsider its behaviour under the combined diffeomorphism plus gauge transformation, denoted $\hat{\varphi}$, that led to the new $SU(2)$ anomaly on $M=\mathbb{C}P^2$. 

How do the pair of WZ terms transform under $\hat{\varphi}$? Recall that $\hat{\varphi}$ is the combination of a complex conjugation diffeomorphism $\varphi$ with a $U(2)$ gauge transformation by $W=\begin{psmallmatrix}
    0 & -1\\
    1 & 0
    \end{psmallmatrix}$. The spin-$U(2)$ connection $A=\sigma^3 a$ defined earlier in this Section, which should now be interpreted as a spin-$SU(2)$ connection due to the massive $U(1)$ component $b$ decoupling, is invariant under $\hat{\varphi}$, and hence so is the field strength $F$. The Pontryagin class $\frac{\text{Tr~} R\wedge R}{8\pi^2}$, being a topological invariant~\cite{novikov1966manifolds}, is invariant under the diffeomorphism $\varphi$ and hence invariant under $\hat{\varphi}$. Finally, given $\varphi$ is locally equivalent to a spatial rotation (in four dimensions), and given also that $\phi$ is an $SU(2)$-singlet, the pseudoscalar $\phi$ is invariant under $\hat{\varphi}$. So both WZ terms in (\ref{eq:WZ}) are invariant under the action of $\hat{\varphi}$.

We already know how the partition function varies under $\hat{\varphi}$ due to the chiral fermion contribution, which is precisely the variation given in Eq. (\ref{eq:phihat_on_Z}). Hence, we conclude that if condition (\ref{eq:condition 2}) is violated, in other words if there is an odd number of fermions with isospins $j\in 4\Z_{\geq 0} + 3/2$, then the effective field theory, which is free of perturbative anomalies by virtue of the effective WZ term, does indeed suffer from a $\Z/2$-valued global anomaly in $\hat{\varphi}$. Up to the effects of the WZ terms, we have arrived at precisely the $SU(2)$ theory defined with spin-$SU(2)$ structure that was introduced by Wang, Wen, and Witten to illustrate the new $SU(2)$ anomaly~\cite{Wang:2018qoy}.

In this way, one can in fact disentangle the effects of perturbative anomalies in the $U(2)$ gauge theory with spin-$U(2)$ structure, and isolate an effective theory that suffers from the new $SU(2)$ anomaly at low energies. But it is important to emphasize that this can only be achieved by including WZ terms (or something similar), which enriches the dynamics of the theory -- for instance, in the gauge we have chosen one must include the effects of a pseudoscalar field $\phi$.
The global anomaly that remains would then have precisely the same physical interpretation as the new $SU(2)$ anomaly; it presents a barrier to defining the theory on non-spin manifolds such as $\mathbb{C}P^2$, at least in the absence of couplings to topological degrees of freedom. This fact that the new $SU(2)$ anomaly, unlike the old one, is in a sense still there in $U(2)$, may also be understood from the perspective of cobordism, as we explain in \S \ref{sec:nospin_bordism}. We remark that a similar trick cannot be performed to restore the old $SU(2)$ global anomaly in the $U(2)$ theory.

It is worth spelling out the fact that, as is the case for the new $SU(2)$ anomaly, this residual global anomaly can always be cancelled by coupling to a tQFT (and considerations of cobordism in \S \ref{sec:nospin_bordism} reveal that there can be no further global anomalies). Unlike the WZ term, such topological degrees of freedom would not alter the dynamics of the theory, but would rather embue the theory with topological order in the deep infrared. We postpone such considerations for future work.

One might distil the various ideas at play in this Section into the following statement: 
\begin{equation}  \nn
    \begin{minipage}{0.8\textwidth}
it is possible to write down a consistent $U(2)$ theory of a single isospin-3/2 fermion, that can be defined on non-spin manifolds using a spin-$U(2)$ structure, if one includes a pair of WZ terms to cancel the perturbative anomalies, and couples to a tQFT to cancel the residual global anomaly.
    \end{minipage}
\end{equation}

\section{Cobordism and the absence of $U(2)$ global anomalies} \label{sec:bordism}

Finally, we discuss the connection between our results and cobordism invariants in five dimensions. Such considerations will also enable us to conclude that there are no further anomalies in the $U(2)$ gauge theories we have considered, defined either with or without a spin structure.

\subsection{Case I: with a spin structure} \label{sec:spin_bordism}

For an $SU(2)$ gauge theory defined on a four-manifold $M$ equipped with spin structure, the original $SU(2)$ anomaly is detected by the bordism group 
\begin{equation}
    \Omega_5^{\text{Spin}}(BSU(2))=\Z/2.\label{eq:SU2bordism}
\end{equation}
There is a corresponding cobordism invariant, namely the $\eta$-invariant, which reduces in this case to a 5d mod 2 index because the fermions are in real representations. Let $\mathcal{I}_{1/2}$ denote this 5d mod 2 index for a single fermion with isospin-$1/2$. For anomalous fermion content, $\mathcal{I}_{1/2}$ is non-vanishing on the mapping torus $M\times S^1$~\cite{Witten:1982fp,Wang:2018qoy}.

When $SU(2)$ is embedded in $U(2)$, a fermion with isospin-$1/2$ is necessarily in a non-trivial representation of $U(1)$ by \eqref{eq: charge constraint}, and thus in a complex representation. Hence, the $\eta$-invariant no longer reduces to a mod 2 index in this case.
But this does not matter in the end, because one may calculate the bordism group directly to find that~\cite{Wan:2019fxh,Davighi:2019rcd}
\begin{equation}
    \Omega_5^{\text{Spin}}(BU(2))=0.\label{eq:U2bordism}
\end{equation} 
Hence, in the case that perturbative anomalies vanish and the $\eta$-invariant becomes a cobordism invariant, there are no cobordism invariants and thus the $\eta$-invariant must be trivial. We therefore deduce that there are no global anomalies in this theory. This is consistent with our explicit calculation in \S \ref{sec:SU2}, which realised the potentially anomalous global $SU(2)$ gauge transformation to be equivalent to a local $U(2)$ gauge transformation.

These statements can be seen from a slightly different perspective. The exponentiated $\eta$-invariant captures both the global and perturbatives anomalies \cite{Witten:2015aba,Yonekura:2016wuc,Witten:2019bou}. In the current case, this can be seen quite explicitly. The vanishing of the fifth bordism group of $BU(2)$ means that any closed spin five-manifold $X$ equipped with a $U(2)$-bundle structure is a boundary of a six-manifold $Y$ with the $U(2)$ and spin structures extended appropriately. The direct relationship between the $\eta$-invariant on such a five-manifold and the anomaly polynomial $I_6$ is then fixed by the Atiyah--Patodi--Singer (APS) index theorem~\cite{atiyah_patodi_singer_1976}
  \begin{equation}
    \label{eq:APS}
    \text{ind}\left(i\slashed{D}\right) = \int_Y I_6 - \eta_X.
  \end{equation}
Whenever the perturbative anomaly vanishes, $\exp(-2\pi i\eta_X)$ becomes trivial on all closed spin five-manifolds and so there can be no additional anomaly.

On the other hand, when the perturbative anomaly doesn't vanish, we can use (\ref{eq:APS}) to compute the $\eta$-invariant explicitly, from the anomaly polynomial $I_6$.
We may choose the closed five-manifold $X$ to be a mapping torus $X=M\times S^1$, whose ends are glued together using a particular gauge transformation $g(x):M \to U(2)$. For any $g(x)$, indeed for any background $U(2)$ connection, this mapping torus is the boundary of a six-manifold $Y=M\times D^2$ to which the $U(2)$ bundle may be extended, where $D^2$ is a hemisphere (topologically a disc) whose equator coincides with the original $S^1$. Note that, importantly, this cannot be done in general for $SU(2)$, or indeed for $SU(2)\times U(1)$, bundles.\footnote{\label{fn:extension} Recall that, for $G=SU(2)$, a 5d mapping torus $X=M\times S^1$ with an $SU(2)$ 1-instanton gauge field on $M$ and periodic spin structure on $S^1$ cannot be extended to any 6-manifold that it bounds. The exponentiated $\eta$-invariant evaluated on this mapping torus equals the (anomalous) phase accrued by the partition function upon doing a constant gauge transformation by $g(x)=-1\in SU(2)$, in the presence of an instanton background.
If such an $SU(2)$ configuration is embedded in $U(2)$, however, we may 
reproduce the transformation by $g=-1$ via
a flat $U(2)$  connection $a=\frac{1}{2}\dd\phi\,{\bf 1}_2$ with non-trivial holonomy, where $\phi\in[0,2\pi)$ parametrises $S^1$. The spin structure  on $S^1$ 
can then be chosen to be
anti-periodic, which {\em can} be extended to a hemisphere $D^2$ that $S^1$ bounds (since it corresponds to the trivial class in $\Omega^{\text{Spin}}_1(pt) \cong \Z/2$). The gauge configuration on $X$ now becomes ${\bf A}= a+A$ which can also be extended to $D^2$. This is because $\dd \phi$ can be realised as the gauge field of a charge-2 Dirac monopole placed at the centre of the hemisphere, restricted to the equator.} 

 We have that
\begin{equation}
  \label{eq:hemisphere}
  \int_{M\times D^2} I_6 = \frac{1}{2}\int_{M\times S^2} \hat{A}(\mathcal{R}) \tr \exp\left(\frac{\mathcal{F}}{2\pi}\right)\bigg|_6,
\end{equation}
where we have expressed the anomaly polynomial explicitly in terms of the $\hat{A}$-genus (sometimes called the `Dirac genus') and the $\UU(2)$ gauge field $\mathcal{F}$. This can be expanded out to give
\begin{equation}
  \label{eq:anom-poly-expanded}
  \int_{M\times D^2} I_6 = \frac{1}{2}\int_{M\times S^2}\left[\frac{1}{24}p_1(\mathcal{R})\tr \frac{\mathcal{F}}{2\pi} + \frac{1}{3!}\tr\left(\frac{\mathcal{F}}{2\pi}\right)^3\right],
\end{equation}
where $p_1(\mathcal{R})$ is the first Pontryagin class of the tangent bundle. Now, $\int_M p_1(\mathcal{R})$ is a multiple of 48 when the (orientable) four-manifold $M$ is spin, due to a signature theorem of Rochlin, so we can ignore the contribution to $\exp(-2\pi i \eta_X)$ coming from the first term in Eq. \eqref{eq:anom-poly-expanded} and focus only on the second term. For a fermion with charge $q$ under the $\U$ part and isospin-$j$ under the $\SU(2)$ part of the gauge group $\UU(2)$, we can write the $\UU(2)$ gauge field $\mathcal{F}$ in terms of the $\U$ gauge field $f$ and the $\SU(2)$ gauge field $F = F^at^a_j$ as\footnote{We use the convention that $f/2\pi$ represents the first Chern class of the gauge field configuration, hence the extra factor of $1/2$.}
\begin{equation}
  \label{eq:U2-field}
  \mathcal{F} = \frac{1}{2}f q{\bf 1}_{2j+1} + F.
\end{equation}
To see the anomaly, we can choose $\mathcal{F}$ such that $f$ has twice the unit magnetic flux through $S^2$ and $F$ is a one-instanton on $M$ as explained in Footnote~\ref{fn:extension}, whence we obtain
\begin{equation}
  \label{eq:anom-poly-eval}
  \int_{M\times D^2}I_6 = \frac{1}{2}q\int_{S^2}\frac{1}{2}\frac{f}{2\pi}\int_M \frac{1}{8\pi^2}\tr F\wedge F = \frac{1}{2}qT(j),
\end{equation}
and thereby conclude that $\exp(-2\pi i \eta_X) = (-1)^{q T(j)}$. Recall that any fermion with isospin $j\in 2\Z_{\geq 0}+1/2$ necessarily has odd charge $q$. We thus arrive at the same physical outcome as in the usual $\SU(2)$ global anomaly, only that it is now the perturbative anomaly that contributes to the $\eta$-invariant (as we saw already in \S \ref{sec:U2}).

\subsection{Case II: without a spin structure} \label{sec:nospin_bordism}

Recall that for the $SU(2)$ gauge theory defined without spin structure the corresponding bordism group is~\cite{Freed:2016rqq,Guo:2017xex,Wan:2018bns}
\begin{equation}
\Omega_5^{\frac{\text{Spin} \times SU(2)}{\Z/2}}=\Z/2 \times \Z/2. \label{eq:spinSU2bordism}
\end{equation}
A possible basis is given by $\mathcal{I}_{1/2}$ and $\mathcal{I}_{3/2}$, the 5d mod 2 indices associated with a single fermion with isospin-$1/2$ or $3/2$ respectively~\cite{Wang:2018qoy}. The former corresponds to the old $SU(2)$ anomaly, and the latter corresponds to the new one.

Now consider the case of a $U(2)$ gauge theory formulated without a spin structure, but rather using a spin-$U(2)$ structure, as was the subject of \S \ref{sec:no spin structure}. In Appendix \ref{app:bordism} we calculate using the Adams spectral sequence that
\begin{equation}
\Omega_5^{\frac{\text{Spin} \times U(2)}{\Z/2}}=\Z/2. \label{eq:spinU2bordism}
\end{equation}
What is the interpretation of this 5d mod 2 cobordism invariant? And does it signify a possible new global anomaly that we have so far missed?

Fermions in either the isospin-$1/2$  or $3/2$ representations must have odd and thus non-vanishing charge under $U(1)$. Thus, it is not clear how to relate the $\eta$-invariant for this theory to a mod 2 index such as $\mathcal{I}_{1/2}$ or $\mathcal{I}_{3/2}$. Moreover, unlike in \S \ref{sec:spin_bordism}, we cannot use the APS index theorem to compute the $\eta$-invariant for an arbitrary closed five-manifold with spin-$U(2)$ structure, because Eq. (\ref{eq:spinU2bordism}) implies that not all such manifolds are bordant to zero. Fortunately, we may follow Ref.~\cite{Wang:2018qoy} in identifying a mod 2 cobordism invariant dual to the generator of \eqref{eq:spinU2bordism} to be
\begin{equation}
    J(Y)=\int_Y w_2(TY)w_3(TY),\label{eq:J}
\end{equation}
where $Y$ is a closed 5-manifold, and $w_{2,3}(TY)$ are Stiefel--Whitney classes. The crucial point is that $J(Y)$ is a mod 2 cobordism invariant of 5-manifolds with no further structure defined.\footnote{Indeed, the fact that the new $SU(2)$ anomaly can be cancelled by the topological Green--Schwartz mechanism, as noted in footnote 7 above, follows essentially from this fact. } Hence, $J(Y)$ is automatically a cobordism invariant of 5-manifolds with spin-$U(2)$ structure, albeit one that can only be detected on non-spin 5 manifolds. For example, 
\begin{equation}
    J\left( \frac{\mathbb{C}P^2 \times S^1}{\Z/2} \right)=1,
\end{equation}
and thus the Dold manifold $(\mathbb{C}P^2 \times S^1)/\Z/2$\footnote{Here the $\Z/2$ acts as complex conjugation on $\mathbb{C}P^2$, and is the antipodal map on $S^1$.} is a suitable generator for the bordism group \eqref{eq:spinU2bordism}. Because $J(Y)$ vanishes trivially on spin manifolds, it does not appear in either \eqref{eq:SU2bordism} or \eqref{eq:U2bordism}.

In Ref.~\cite{Wang:2018qoy}, the cobordism invariant $J(Y)$ was identified, for any five-manifold with spin-$SU(2)$ structure, with the mod 2 index $\mathcal{I}_{3/2}$, and thus with the new $SU(2)$ anomaly, since the Dold manifold corresponds precisely to the action of the diffeomorphism plus gauge transformation $\hat{\varphi}$ on $\mathbb{C}P^2$. Since the action of $\hat{\varphi}$ on the corresponding $U(2)$ theory is equivalent, at the level of the partition function, to a local $U(2)$ transformation as described in \S \ref{sec:no spin structure}, the potential global anomaly corresponding to this cobordism invariant necessarily vanishes by perturbative anomaly cancellation. That said, as we saw in \S \ref{sec:disentangle}, by including WZ terms to cancel the perturbative anomalies in the low energy effective theory, it is possible to reveal a low-energy theory which does indeed suffer from this `new $U(2)$ anomaly', which corresponds to the $\Z/2$ in (\ref{eq:spinU2bordism}). Since there are no other independent cobordism invariants, we conclude that there are no other possible global anomalies in the $U(2)$ gauge theory defined using a spin-$U(2)$ structure.

\subsection*{Acknowledgments}
We thank Ben Gripaios and David Tong for discussions, and Pietro Benetti Genolini for reading the manuscript. We also thank the anonymous referee for their very interesting suggestions. JD is supported by the STFC consolidated grant ST/P000681/1. NL is supported by the DPST Scholarship from the Thai Government.

\appendix

\section{Spin-\texorpdfstring{$U(2)$}{U(2)} bordism} \label{app:bordism}

In this Appendix we calculate the bordism group $\Omega_5^{\frac{\text{Spin} \times U(2)}{\Z/2}}(\text{pt})$, using the Adams spectral sequence. For a guide to using the Adams sequence to compute bordism groups, we recommend Ref.~\cite{beaudry2018guide}.

When there is no odd-torsion involved, the bordism group $\Omega^G_{t-s}(\text{pt})$ can be evaluated via the Adams spectral sequence
\begin{equation}
  \label{eq:1}
  \ext^{s,t}_{\A}(H^\bullet(MTG),\Z/2) \Rightarrow \Omega^G_{t-s}(\text{pt}),
\end{equation}
where $\A$ is the Steenrod algebra and $MTG$ is the Madsen--Tillmann spectrum defined in terms of the Thom spectrum by $MTG = \text{Thom}(BG,-V)$, with $V$ a stable bundle of virtual dimension 0 pulled back from the tautological stable bundle over $BO$ by $BG\rightarrow BO$. In our case, $MTG$ can be written as
\begin{equation}
  \label{eq:2}
  MTG = M\Spin \wedge X_G,
\end{equation}
with $X_G$ a Thom spectrum to be determined. For $t-s<8$, this simplifies the Adams spectral sequence above to
\begin{equation}
  \label{eq: main Adams ss}
  \ext^{s,t}_{\A_1}(H^\bullet(X_G),\Z/2)\Rightarrow \Omega^G_{t-s}(\text{pt}),
\end{equation}
by the Anderson-Brown-Peterson theorem. Here $\A_1$ denotes the subalgebra of $\A$ generated by the Steenrod operations $\sq^1$ and $\sq^2$. To make the presentation clearer, we will write  $\UU_n$ and $\SO_n$ for $\UU(n)$ and $\SO(n)$ in the rest of this Appendix.

\subsection*{Calculation of \texorpdfstring{$X_G$}{XG}}
\label{sec: XG calc}

 We will now show that the Thom spectrum $X_G$ when $G = (\Spin\times\UU_2)/\Z/2$ is  given by $X_G = \Sigma^{-5}M\SO_3\wedge M\UU_1$. We follow the calculation of related examples in Refs.~\cite{Wan:2018bns,Wan:2019fxh}, whose method was based on Ref.~\cite{Freed:2016rqq}.

The fibration $\fib{\Z/2}{G}{\SO\times \SO_3\times \UU_1}$ gives rise to the following fibration sequence of classifying spaces
\begin{equation}
  \label{eq: BG fib}
  BG \xrightarrow{(f,f^\prime,f^\dprime)} B\SO\times B\SO_3\times B\UU_1 \xrightarrow{w_2+w_2^\prime+w_2^\dprime} K(\Z/2,2),
\end{equation}
where $w_2\in H^2(B\SO), w_2^\prime\in H^2(B\SO_3)$, and $w_2^\dprime\in H^2(B\UU_1)$ are the second Stiefel--Whitney classes for $B\SO$, $B\SO_3$, and $B\UU_1$, respectively. The fibration sequence \eqref{eq: BG fib} arises as a Puppe sequence, so the composite map
\begin{equation}
  w_2\circ f + w_2^\prime\circ f^\prime + w_2^\dprime\circ f^\dprime: BG \rightarrow K(\Z/2,2)\nn
\end{equation}
is null-homotopic. Moreover, since these classes are valued modulo $2$, this is equivalent to saying that the map $w_2\circ f$ is homotopy equivalent to $w_2^\prime \circ f^\prime + w_2^\dprime \circ f^\dprime$. Therefore, the following diagram 
\begin{equation}
  \label{eq: prim pb}
  \begin{tikzcd}
    BG \arrow[r,"{(f^\prime,f^\dprime)}"] \arrow[d, "f"] & B\SO_3\times B\UU_1 \arrow[d, "w_2^\prime+w_2^\dprime"]\\
    B\SO \arrow[r, "w_2"] & K(\Z/2,2)
  \end{tikzcd}
\end{equation}
is a homotopy pullback square, which we also use to define the map $V:\;BG \xrightarrow{f} B\SO\incl B\O$. 

Equivalently,  $BG$ fits into the homotopy pullback
\begin{equation}
  \label{eq: sec pb}
\begin{tikzcd}
BG \arrow[r] \arrow[d, "{(f,f^\prime,f^\dprime)}"]
& B\Spin \arrow[d, "g"] \\
B\SO\times B\SO_3\times B\UU_1 \arrow[r, "h"]
& B\SO \arrow[r,"w_2"] & K(\Z/2,2)
\end{tikzcd}
\end{equation}
where $w_2\circ g$ is null-homotopic and $h$ is to be determined. This can be seen by finding a suitable map $h$, as follows. Since $BG$ fits into the homotopy pullback \eqref{eq: prim pb}, we can think of its element as a triplet of vector bundles $(V,V_3,V_2)\in B\SO\times B\SO_3\times B\UU_1$, such that $w_2(V) = w_2(V_3)+w_2(V_2)$. We take the map $h$  from $BG$ to $B\SO$ to be
\begin{equation}
  \label{eq:11}
  (V,V_3,V_2) \mapsto  V + V_3 + V_2 -5,
\end{equation}
which sends three bundles into a stable $\SO$-bundle of virtual dimension 0. Using the Whitney product formula, the second Stiefel--Whitney class of the virtual bundle $V+V_3+V_2-5$ is given by
\begin{equation}
  \label{eq:14}
  w_2(V+V_3+V_2-5) = w_2(V) + w_2(V_3) + w_2(V_2) = 0
\end{equation}
where we obtain the last equality using the pullback square \eqref{eq: prim pb}. Therefore, the stable $\SO$-bundle $V+V_3 + V_2-5$ can be lifted to a stable spin bundle, denoted by $W$, establishing the existence of a homotopy pullback \eqref{eq: sec pb}.

Therefore, the map  $-V: BG\rightarrow B\SO$ is homotopy equivalent to the map  $-W + V_3 + V_2 -5$  from $B\Spin\times B\SO_3\times B\UU_2$ into $B\SO$, giving rise to the identification of the Thom spectrum $MTG= \th(BG;-V)$ with
\begin{equation}
  \label{eq: thom iden}
\th (B\Spin\times B\SO_3\times B\UU_1; -W + V_3+V_2 -5) = \Sigma^{-5} M\Spin\wedge M\SO_3\wedge M\UU_1.
\end{equation}

\subsection*{\texorpdfstring{$\A_1$}{A1}-module structure of \texorpdfstring{$H^\bullet(X_G)$}{H(XG)} and Adams spectral sequence}
\label{sec: mod Adams}

\begin{figure}[!htb]
\centering
      \includegraphics[width=0.5\textwidth]{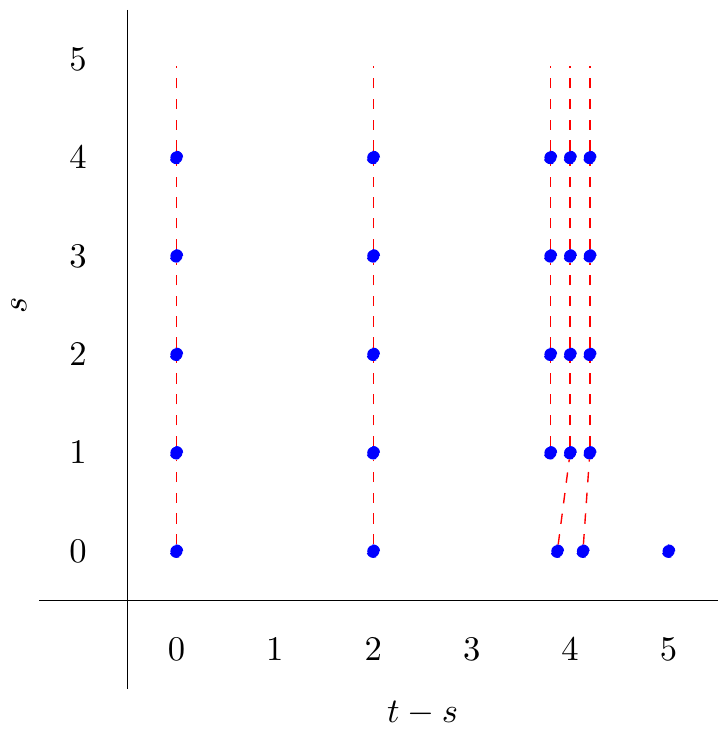}
\caption{The $E_2$ page of the Adams spectral sequence (\ref{eq: main Adams ss}), from which one can read off the bordism groups $\Omega_{d \leq 5}^{\frac{\text{Spin} \times U(2)}{\Z/2}}(\text{pt})$. \label{fig: E2 ASS}}
\end{figure}

We will now work out the $\A_1$-module structure of the spectrum $X_G$. Recall that 
\begin{equation}
H^\bullet(B\SO_3) \cong \Z/2[w_2^\prime,w_3^\prime]\quad \text{and} \quad H^\bullet(B\UU_1) \cong \Z/2[w_2^\dprime],   
\end{equation}
where $w_2^\prime,w_3^\prime$ are the Stiefel--Whitney classes,  with $w_2^\dprime$ being the first Chern class modulo 2, which coincides with the second Stiefel--Whitney class. By the Thom isomorphism, we have the identifications
\begin{equation}
  \label{eq: Thom iden}
  H^\bullet(M\SO_3) \cong \Z/2[w_2^\prime,w_3^\prime]\{U\} \quad \text{and} \quad
  H^\bullet(M\UU_1) \cong \Z/2[w_2^\dprime]\{V\},
\end{equation}
where the Thom classes $U$ and $V$ are in $H^3(M\SO_3)$ and $H^2(M\UU_1)$ respectively. The K\"{u}nneth theorem for the cohomology ring of a Thom space  implies that 
\begin{align}
  \label{eq: Kunneth}
  H^\bullet(\Sigma^{-3}M\SO_3\wedge \Sigma^{-2}M\UU_1) &\cong \Sigma^{-5} H^\bullet(M\SO_3)\otimes H^\bullet (M\UU_1)\nn\\ 
  &\cong \Sigma^{-5}\Z/2[w_2^\prime,w_3^\prime,w_2^\dprime]\{UV\}.
\end{align}
Using the relations between Thom classes, the Steenrod squares, and the Stiefel--Whitney classes, we find that the $\A_1$-module structure of $H^\bullet(X_G)$ up to degree 5 can be expressed as the cell diagram shown in Fig. \ref{fig: SO3U1 mod}, with the corresponding Adams chart for $\ext^{s,t}_{\A_1}(H^\bullet(X_G),\Z/2)$ shown in Fig.~\ref{fig: E2 ASS}. 
 In the Adams chart, each dot corresponds to a $\Z/2$ generator.  A line joining two generators $\alpha_s$ and $\alpha_{s+1}$ of the same $t-s$ but with $\Delta s = 1$ means that the generator $\alpha_{s+1}$ is given by $\alpha_{s+1}= h_0 \alpha_s$, where $h_0$ is the generator of $\ext^{1,1}_{\A_1}(\Z/2,\Z/2)$.
\begin{figure}[!htb]
\centering
\begin{tikzpicture}[scale=.5] 
  \fill (0, 0) circle (3pt) node[anchor=west]  {{\tiny $UV$}};
  \fill (0, 2) circle (3pt) node[anchor=east] {};
  \sqtwoL (0, 0, black);

  \fill (0,3) circle (3pt) node[anchor=east] {};
  \fill (0,4) circle (3pt) node[anchor=west] {};
  \sqone (0,2, black);
  \fill (0,5) circle (3pt) node[anchor=south] {};
  \sqtwoL (0,3,black);
  \sqone (0,4,black);

  \fill (2,2) circle (3pt) node[anchor=north] {{\tiny $w_2^\dprime UV$}};
  \sqtwoCL (2,2,black);

  \fill (2,4) circle (3pt) node[anchor=north] {{\tiny $w_2^{\dprime 2}UV$}};
  \fill (2,6) circle (3pt) node[anchor=east] {};
  \sqtwoL (2,4,black);
  \sqone (2,6,black);
  \fill (2,7) circle (3pt) node[anchor=east] {};
  \sqtwoL (2,7,black);
  \fill (2,9) circle (3pt) node[anchor=east] {};

  \fill (4,6) circle (3pt) node[anchor=east] {};
  \fill (2,8) circle (3pt) node[anchor=west] {};
  \sqtwoCL (4,6,black);
  \sqone (2,8,black);

  \fill (6,4) circle (3pt) node[anchor=north] {{\tiny $w_2^{\prime 2} UV$}};
  \fill (6,6) circle (3pt) node[anchor=east] {};
  \sqtwoL (6,4,black);
  \sqone (6,6,black);
  \fill (6,7) circle (3pt) node[anchor=east] {};
  \sqtwoL (6,7,black);
  \fill (6,9) circle (3pt) node[anchor=east] {};
  \fill (6,8) circle (3pt) node[anchor=east] {};
  \sqone (6,8,black);
  \fill (8,6) circle (3pt) node[anchor=east] {};
  \sqtwoCL (8,6,black);
  
  \fill (8,8) circle (3pt) node[anchor=east] {};
  \sqone (8,8,black);
  \fill (8,9) circle (3pt) node[anchor=east] {};
  \sqtwoL (8,9,black);
  \fill (8,11) circle (3pt) node[anchor=east] {};

  \fill (10,5) circle (3pt) node[anchor=north] {{\tiny $w_2^\prime w_3^\prime UV$}};
  \sqone (10,5,black);
  \sqtwoR (10,5, black);
  \fill (10,6) circle (3pt) node[anchor=east] {};
  \fill (10,7) circle (3pt) node[anchor=west] {};
  \sqtwoCL (10,7,black);
  \sqone (10,7,black);
  \fill (10,8) circle (3pt) node[anchor=west] {};
  \sqtwoCL (10,6,black);
  \fill (8,10) circle (3pt) node[anchor=west] {};
  \sqtwoCL (10,8,black);
  \sqone (8,10,black);
\end{tikzpicture}
\caption{The $\A_1$-module structure for $\Z/2[w_2^\prime,w_3^\prime,w_2^\dprime]\{UV\}$, up to degree ten.\label{fig: SO3U1 mod}}
\end{figure}

In the range of our interest ($t-s<6$), the entries are too sparse and all the differentials are trivial, apart from a possible non-trivial differential $d_r$ from the entry $(s,t-s) = (0,5)$ to the entries $(s,t-s) = (r,4)$. However, using the fact that $d_r$ commutes with $h_0$, it can be shown that these differentials are trivial, too. Therefore, the Adams spectral sequence collapses already at the $E_2$ page for $t-s < 6$. 

Finally, the rule for extracting the bordism groups can be roughly summarised as follows: an $h_0$-tower containing $m$ dots gives a factor of $\Z/2^m$, and an infinite $h_0$-tower gives a factor of $\Z$. With this rule, the bordism groups of degree lower than six can be read off from the chart in Fig.~\ref{fig: E2 ASS} to be
\begin{equation}
  \label{eq:bordism}
  \Omega^G_0 = \Z, \quad \Omega^G_1 = 0, \quad \Omega^G_2 = \Z,\quad \Omega^G_3 = 0,\quad \Omega^G_4 = \Z^3,
\end{equation}
and, crucially for us, 
\begin{equation}
    \Omega^G_5 = \Z/2.
\end{equation}

\bibliographystyle{JHEP}
\bibliography{references}
\end{document}